\documentclass[preprint,5p,twocolumn]{elsarticle}

\usepackage{graphicx}
\usepackage{subcaption}
\usepackage{lipsum}
\usepackage{booktabs}
\usepackage{tabularx}
\usepackage{tikz}
\usepackage{pgfplots}
\usetikzlibrary{positioning,shapes,trees,arrows,graphs,decorations,calc}
\usepackage{url}

\usepackage{amssymb}
\usepackage{amsmath}
\usepackage{hyperref}
\usepackage{xcolor}

\usepackage{placeins} 
\usepackage[normalem]{ulem}
\usepackage{xspace}

\newcommand{\eq}[1]{eq.~\eqref{eq:#1}}
\newcommand{\eqs}[2]{eqs.~\eqref{eq:#1} and \eqref{eq:#2}}
\renewcommand{\sec}[1]{sec.~\ref{sec:#1}}

\newcommand{\fig}[1]{Fig.~\ref{fig:#1}}

\newcommand{\apps}[2]{\ref{app:#1} and \ref{app:#2}}

\newcommand{\ord}[1]{\mathcal{O}(#1)}

\newcommand{\df}{\mathrm{d}}

\newcommand{\Tau}{\mathcal{T}}

\newcommand{\GeV}{\,\mathrm{GeV}}
\newcommand{\TeV}{\,\mathrm{TeV}}

\newcommand{\nn}{\nonumber}

\newcommand{\cP}{\mathcal{P}}

\newcommand{\cut}{\mathrm{cut}}

\newcommand{\NLO}{\mathrm{NLO}}

\newcommand{\one}{{(1)}}

\newcommand{\dsigMC}{\df\sigma^\textsc{mc}}
\newcommand{\dsigtildeMCz}{\widetilde{\df\sigma^\textsc{mc}_0}}

\newcommand{\geneva}{\textsc{Geneva}\xspace}

\newcommand{\powheg}{\textsc{Powheg}\xspace}

\newcommand{\pythia}{\textsc{Pythia}\xspace}
\newcommand{\pythiaEight}{\textsc{Pythia8}\xspace}

\newcommand{\openloopsTwo}{\textsc{OpenLoops2}\xspace}

\renewcommand{\matrix}{\textsc{Matrix}\xspace}
\newcommand{\munich}{\textsc{Munich}\xspace}

\newcommand{\T}{\mathrm{T}}

\biboptions{comma,compress}

\newcounter{bla}

\newcommand{\rescalethreeplots}{0.32\textwidth}
\newcommand{\hspacebetweenthreeplots}{1em}

\newcommand{\spaceabovefigurecaption}{-1ex}

\makeatletter
\g@addto@macro\bfseries{\boldmath}
\makeatother

\begin{document}

\begin{frontmatter}

\title{Next-to-next-to-leading order event generation for $Z$-boson
   pair\\ production  matched to parton shower}

\author[label1]{Simone Alioli}
\author[label1]{Alessandro Broggio}
\author[label1]{Alessandro Gavardi}
\author[label1]{Stefan Kallweit}
\author[label1]{\\ Matthew A.~Lim}
\author[label1]{Riccardo Nagar}
\author[label1]{Davide Napoletano}

\cortext[author] {Corresponding author.\\\textit{E-mail address:}
  simone.alioli@unimib.it}

\address[label1]{Universit\`a degli Studi di Milano-Bicocca \& INFN
  Sezione di Milano-Bicocca, Piazza della Scienza 3, Milano 20126,
  Italy}

\date{\today}

\begin{abstract}
  We present a novel next-to-next-to-leading order (NNLO) QCD
  calculation matched to parton shower for the production of a pair of
  $Z$ bosons decaying to four massless leptons,
  $p p \to \ell^+ \ell^- \ell'^+ \ell'^- + X$, at the LHC.  Spin
  correlations, interferences and off-shell effects are included
  throughout.  Our result is based on the resummed beam-thrust
  spectrum, which we evaluate at next-to-next-to-leading-logarithmic
  (NNLL$^\prime_{\Tau_0}$) accuracy for the first time for this
  process, and makes use of the \geneva Monte Carlo framework for the
  matching to \pythiaEight shower and hadronisation models. We compare
  our predictions with data from the ATLAS and CMS experiments at
  $13 \TeV$, finding a good agreement.

\end{abstract}

\end{frontmatter}

\section{Introduction}
\label{sec:intro}

Diboson production at the LHC is of paramount importance to the
precision study of electroweak (EW) physics  in the Standard Model~(SM)
and beyond, as it directly probes the non-Abelian EW couplings.
The four-lepton channel that proceeds predominantly
through intermediate $Z$-boson pairs is of particular importance due
to its very clean signature.  Consequently, several cross-section
measurements and studies of anomalous couplings were carried out by
both the ATLAS and CMS collaborations at
7\,TeV~\cite{Aad:2011xj,Aad:2012awa,CMS:2012bw,Chatrchyan:2012sga},
8\,TeV~\cite{Aad:2015rka,Aaboud:2016urj,Chatrchyan:2013oev,CMS:2014xja,Khachatryan:2015pba}
and
13\,TeV~\cite{Aad:2015zqe,Aaboud:2017rwm,Aaboud:2019lxo,Aaboud:2019lgy,Khachatryan:2016txa,Sirunyan:2017zjc,Sirunyan:2020pub}. Moreover,
analyses of the four-lepton channel have helped to constrain the width
and couplings of the Higgs
boson~\cite{Aad:2012tfa,Aad:2014eva,Aad:2015xua,Aaboud:2018puo,ATLAS:2020wny,Chatrchyan:2012ufa,Chatrchyan:2013mxa,Khachatryan:2014iha,Khachatryan:2015mma,Khachatryan:2016ctc,Sirunyan:2017tqd,Sirunyan:2019twz}.
In addition, given that $Z'$ resonances are common features in models
of new physics, measurements of $Z$ pair production can help to set
limits on Beyond the SM scenarios.

Theoretical predictions to $ZZ$ production have been known at NLO in QCD
for some time~\cite{Mele:1990bq,Ohnemus:1990za,Ohnemus:1994ff,Dixon:1999di,Campbell:1999ah,Dixon:1998py}.
These were followed by calculations of NLO EW corrections, both for on-shell bosons~\cite{Accomando:2004de,Bierweiler:2013dja,Baglio:2013toa} and for fully leptonic final states~\cite{Biedermann:2016yvs,Biedermann:2016lvg}, and more recently  combinations of NLO QCD and EW contributions have appeared~\cite{Kallweit:2017khh,Chiesa:2018lcs}. The current state-of-the-art at fixed order in perturbation theory is NNLO~\cite{Cascioli:2014yka,Heinrich:2017bvg,Grazzini:2015hta,Kallweit:2018nyv} in QCD for the process \mbox{$q\bar{q}\to ZZ \to 4\ell$}, combined with NLO EW effects~\cite{Kallweit:2019zez}.
NLO corrections (at $\ord{\alpha_s^3}$) to the loop-induced process \mbox{$gg\to ZZ \to 4\ell$} are also available~\cite{Caola:2015psa,Caola:2016trd,Grazzini:2018owa,Grazzini:2021iae}.

In order to take advantage of the increasing precision of experimental
data, it is useful to provide predictions in the form of fully
exclusive Monte Carlo (MC) event generators. These allow
hadron-level events to be produced that can be directly
interfaced to detector simulations and experimental analyses.

For the \mbox{$q\bar{q}\to ZZ \to 4\ell$} channel, the state-of-the-art in this regard is still matching NLO QCD correction to parton shower (NLOPS), as implemented in \powheg~\cite{Melia:2011tj,Nason:2013ydw}, also including NLO EW effects~\cite{Chiesa:2020ttl}. 
For the \mbox{$gg\to ZZ \to 4\ell$} channel NLOPS predictions were presented in Refs.~\cite{Alioli:2016xab,Alioli:2021wpn}.

There
has, however, been significant progress in the matching of NNLO calculations to parton showers~(NNLOPS) in recent years, with four major approaches to the problem~\cite{Hamilton:2013fea,Alioli:2012fc,Alioli:2013hqa,Alioli:2015toa,Hoeche:2014aia,Hoche:2014dla,Monni:2019whf,Monni:2020nks}. In this Letter, we consider the \geneva framework developed in~\cite{Alioli:2012fc,Alioli:2015toa,Alioli:2021qbf}, which has been successfully applied to the Drell--Yan~\cite{Alioli:2015toa} and Higgsstrahlung~\cite{Alioli:2019qzz} processes, as well as diphoton production~\cite{Alioli:2020qrd} and hadronic Higgs decays~\cite{Alioli:2020fzf}.

Currently, the only available NNLOPS calculation featuring two massive bosons
in the final state is for the process $pp\to W^+W^- \to \ell \nu \ell' \nu'$~\cite{Re:2018vac} 
via the \textsc{MiNLO}$^\prime$ method~\cite{Hamilton:2013fea},
which made use of a differential reweighting to the \matrix predictions of Ref.~\cite{Grazzini:2016ctr} to achieve NNLO accuracy. Na\"ively, the complexity of the final state would require a nine-dimensional reweighting, something unfeasible in
practical terms. In Ref.~\cite{Re:2018vac}, the authors were able to circumvent this limitation by rewriting the
differential cross section in terms of angular coefficients, which they used
to reweight each angular contribution separately. The dimensionality of the
reweighting procedure was thus reduced to just three. This relied, however, on the
approximation that the vector bosons are close to being on-shell, and so
cannot be easily applied to the case of $pp \to Z/\gamma^* Z/\gamma^* \to
4\ell  $, given the non-negligible contribution  from photon exchange.

Since the \geneva method does not need a multi-differential reweighting to
reach NNLO accuracy, we are able to include the full off-shell effects and
deliver the first NNLOPS calculation for $ZZ$ production.\footnote{We note that  the methods of Refs.~\cite{Hoeche:2014aia,Hoche:2014dla} and Refs.~\cite{Monni:2019whf,Monni:2020nks} also do not require a reweighting procedure to reach NNLO accuracy.}

The process $pp\to \ell^+\ell^- \ell'^+ \ell'^- + X$ features
contributions from channels with very different resonance
structures.
In order to increase efficiency in event generation in
such a situation, it is necessary to make use of a phase space
generator which samples these channels separately.
To this end, we have
built 
an interface between \geneva and the multi-channel integrator \munich~\cite{munich} which is completely general and allows for the integration of any SM process.

This Letter is organised as follows: we discuss the process definition and  the relevant calculation in \sec{proc}, present our results and the comparison to LHC data in \sec{results} and draw our conclusions in \sec{conc}. Additional details concerning the calculation are provided in  \apps{power_corrections}{profile_functions}.

\section{Process definition}
\label{sec:proc}

In this Letter, we consider the process \mbox{$pp\to \ell^+\ell^- \ell'^+ \ell'^- + X$}, including off-shell effects, $Z/\gamma^*$ interference and spin correlations.
The derivation of the \geneva formulae has been presented in several papers, see \textit{e.g.}~\cite{Alioli:2015toa,Alioli:2019qzz}. Here we content ourselves with presenting the final results for the differential weights of the $0$-,$1$- and $2$- jet partonic cross sections.

These are defined in such a way that they correspond to  physical and IR-finite events at a given perturbative accuracy, with the condition that IR singularities cancel on an event-by-event basis. The \geneva method  achieves this by mapping IR-divergent final states with $M$ partons into IR-finite final states with $N$ jets, with $M \geq N$. 
Events are classified according to the value of $N$-jet resolution variables $\Tau_N$ which partition the phase space into different regions according to the number of resolved emissions.
In particular, the \geneva Monte Carlo cross section $\dsigMC_N$ receives
contributions from both $N$-parton events and $M$-parton events where the additional emission(s) are below the resolution cut $\Tau_N^\cut$ used to separate resolved and unresolved emissions.
The unphysical dependence on the boundaries of this partitioning procedure is removed by requiring that the resolution parameters are resummed at high enough accuracy.
For the production of a pair of $Z$ bosons at NNLO accuracy, we need to introduce the $0$-jettiness resolution variables $\Tau_0$   and $1$-jettiness $\Tau_1$ defined as 
\begin{align} \label{eq:TauNdef}
\Tau_N = \sum_k \min \Bigl\{ \hat q_a \cdot p_k, \hat q_b \cdot p_k, \hat q_1 \cdot p_k, \ldots , \hat q_N \cdot p_k \Bigr\}
\,,\end{align}
with $N=0$ or $1$,  $q_a, q_b$ representing the beam directions and $q_k$ any final state massless direction that minimise $\Tau_N$. These separate the $0$- and $1$-jet exclusive  cross sections from the $2$-jet inclusive one.

We find that
\begin{align} \label{eq:0masterful}
\frac{\dsigMC_0}{\df\Phi_0}(\Tau_0^\cut)
&= \frac{\df\sigma^{\rm NNLL'}}{\df\Phi_0}(\Tau_0^\cut)
- \biggl[\frac{\df\sigma^{\rm NNLL'}}{\df\Phi_0}(\Tau_0^\cut) \biggr]_{\rm NNLO_0}
\nn \\ & \quad
+ (B_0 + V_0 + W_0)(\Phi_0)
\nn \\ & \quad
+ \int \! \frac{\df \Phi_1}{\df \Phi_0}\, (B_1 + V_1)(\Phi_1)\, \theta[\Tau_0(\Phi_1) < \Tau_0^\cut]
\nn \\ & \quad
+ \int \! \frac{\df \Phi_2}{\df \Phi_0}\, B_2 (\Phi_2)\, \theta[\Tau_0(\Phi_2) < \Tau_0^\cut]
\,,
\end{align}
\begin{align}
\label{eq:1masterful}
&\frac{\dsigMC_1}{\df\Phi_1} (\Tau_0 > \Tau_0^\cut; \Tau_1^\cut)
\nn \\ & \quad
= \Bigg\{ \frac{\df\sigma^{\rm NNLL'}}{\df\Phi_0\df\Tau_0}\cP(\Phi_1)
+ (B_1 + V_1^C)(\Phi_1)
\nn \\ & \quad \quad \quad \quad
- \biggl[\frac{\df\sigma^{\rm NNLL'}}{\df\Phi_0\df\Tau_0}\cP(\Phi_1)\,\biggr]_{\NLO_1}\Bigg\}
\nn \\ & \quad
\times U_1(\Phi_1, \Tau_1^\cut)\, \theta(\Tau_0 > \Tau_0^\cut)
\nn \\ & \quad
+\int\!\biggl[\frac{\df\Phi_2}{\df\Phi_1^\Tau}\,B_2(\Phi_2)\, \theta[\Tau_0(\Phi_2) > \Tau_0^\cut]\,\theta(\Tau_1 < \Tau_1^\cut)
\nn \\ & \quad \quad \quad \quad
- \frac{\df\Phi_2}{\df \Phi_1^C}\, C_2(\Phi_2)\, \theta(\Tau_0 > \Tau_0^\cut) \biggr]
\nn \\ & \quad
- B_1(\Phi_1)\, U_1^\one(\Phi_1, \Tau_1^\cut)\, \theta(\Tau_0 > \Tau_0^\cut)
\,,
\end{align}
\begin{align}
\label{eq:1NS}
\frac{\dsigMC_1}{\df\Phi_1}(\Tau_0 & \leq \Tau_0^\cut)   \\ = \nn &\ \overline{\Theta}^{\mathrm{FKS}}_{\mathrm{map}}(\Phi_1)  \, (B_1+V_1)\, (\Phi_1)\,\theta(\Tau_0<\Tau^{\mathrm{cut}}_0) \,,
\end{align}
\begin{align}
\label{eq:2masterful}
&\frac{\dsigMC_{\geq 2}}{\df\Phi_2} (\Tau_0 > \Tau_0^\cut,
\Tau_1>\Tau_1^\cut) 
\nn \\  & \quad
=\Bigg\{ \frac{\df\sigma^{\rm NNLL'}}{\df\Phi_0\df\Tau_0}\cP(\Phi_1)
+ (B_1 + V_1^C)(\Phi_1)
\nn \\ & \quad \quad \quad \quad
- \biggl[\frac{\df\sigma^{\rm NNLL'}}{\df\Phi_0\df\Tau_0}\cP(\Phi_1)\,\biggr]_{\NLO_1}\Bigg\}
\nn \\ & \quad
\times U_1'(\Phi_1, \Tau_1) \theta(\Tau_0\!\! >\!\! \Tau_0^\cut) \Big\vert_{\Phi_1 = \Phi_1^\Tau\!\!(\Phi_2)}\!\! \!\! \cP(\Phi_2) \, \theta(\Tau_1\!\! >\!\! \Tau_1^\cut)
\nn \\ & \quad
+\bigl\{ B_2(\Phi_2)\,[1 - \Theta^\Tau(\Phi_2)\,\theta(\Tau_1 < \Tau_1^\cut)]
\nn \\ & \quad \quad  \quad
- B_1(\Phi_1^\Tau)\,U_1^{\one\prime}(\Phi_1^\Tau, \Tau_1)\,\cP(\Phi_2)\,\theta(\Tau_1 > \Tau_1^\cut)
\bigr\}
\nn \\ & \quad
\times \theta[\Tau_0(\Phi_2) > \Tau_0^\cut]
\,,\end{align}

\begin{align}
\label{eq:2NS}
&\frac{\dsigMC_{\geq 2}}{\df\Phi_2}  (\Tau_0 > \Tau_0^\cut, \Tau_1 \le \Tau_1^\cut)
 \\ & \quad
=   B_2(\Phi_2)\, \overline{\Theta}^\Tau(\Phi_2)\,\theta(\Tau_1 < \Tau_1^\cut)\, \theta\left(\Tau_0(\Phi_2) > \Tau_0^\cut\right) \, ,\nn
\end{align}
where the $B_j$, $V_j$ and $W_j$ are the $0$-, $1$- and $2$-loop
matrix elements for $j$ QCD partons in the final state.

We have introduced the shorthand notation
\begin{align}
\label{eq:dPhiRatio}
 \frac{\df \Phi_{M}}{\df \Phi_N^{\cal O}}  = \df \Phi_{M} \, \delta[ \Phi_N - \Phi^{\cal O}_N(\Phi_M) ] \,\Theta^{\cal O}(\Phi_N)
\end{align}
to indicate that the integration over a region of the $M$-body phase space
is performed while keeping the $N$-body phase space and the value of some specific
observable $\cal O$ fixed, with $N \leq M$. The $\Theta^{\cal
  O}(\Phi_N)$ term in the previous equation limits the integration to
the phase space points included in the singular contribution for the
given observable $\cal O$.  For example, when generating $1$-body
events we use
\begin{equation} \label{eq:Phi1TauProj}
\frac{\df\Phi_2}{\df\Phi_1^\Tau} \equiv \df\Phi_2\,\delta[\Phi_1 - \Phi^\Tau_1(\Phi_2)]\,\Theta^\Tau(\Phi_2)
\,,\end{equation}
where the map used by the $1 \to 2$ splitting has been constructed
to preserve $\Tau_0$, i.e.
\begin{equation} \label{eq:Tau0map}
\Tau_0(\Phi_1^\Tau(\Phi_2)) = \Tau_0(\Phi_2)
\,,\end{equation}
and $\Theta^\Tau(\Phi_2)$ defines the projectable region of $\Phi_2$ which can be reached starting from a point in $\Phi_1$ with a specific value of $\Tau_0$.
The use of a $\Tau_0$-preserving mapping is necessary to ensure that
the pointwise singular $\Tau_0$ dependence is alike among all terms in
\eqs{1masterful}{2masterful} and that the cancellation of said singular
terms is guaranteed on an event-by-event basis.

The expressions in \eqs{1NS}{2NS} encode the nonsingular contributions
to the $1$- and $2$-jet rates which arise from non-projectable
configurations below the corresponding cut. This is highlighted by the
appearance of the complementary $\Theta$ functions, $\overline{\Theta}^{\cal O}$, which account for
any configuration that is not projectable either because it would
result in an invalid underlying-Born flavour structure or because it
does not satisfy the $\Tau_0$-preserving mapping (see also
\cite{Alioli:2019qzz}).

The term $V_1^C$ denotes the soft-virtual contribution of a standard NLO local subtraction (in our implementation, we follow
the FKS subtraction as detailed in \cite{Frixione:2007vw}). We have that
\begin{align} \label{eq:FOFKS}
  V_1^C(\Phi_1) = V_1(\Phi_1)+\int\frac{\df\Phi_2}{\df \Phi_1^C}C_2(\Phi_2)\,,
\end{align}
 with $C_2$ a singular approximation of $B_2$: in practice we use the
 subtraction counterterms which we integrate over the radiation variables
 $\df\Phi_2 / \df \Phi_1^C$  using the singular limit $C$ of the phase space
 mapping. $U_1$ is an NLL Sudakov factor which resums large logarithms of $\Tau_1$ and $U_1'$ its derivative with respect to $\Tau_1$.

The term $\mathcal{P}(\Phi_{N+1})$
represents a normalised splitting probability which
serves to extend the differential dependence of the resummed terms
from the $N$-jet to the $(N\!+\!1)$-jet phase space. For example, in
\eq{1masterful}, the term $\cP(\Phi_1)$ makes the resummed spectrum in
the first term (which is naturally differential in the $\Phi_0$
variables and $\Tau_0$) differential also in the additional two
variables needed to cover the full $\Phi_1$ phase space. These
splitting probabilities are normalised, i.e. they satisfy
\begin{align}
\label{eq:cPnorm}
\int \! \frac{\df\Phi_{N+1}}{\df \Phi_{N} \df \Tau_N} \, \cP(\Phi_{N+1}) = 1
\,.\end{align}
The two extra variables are chosen to be an energy ratio $z$ and an
azimuthal angle $\phi$. The functional forms of the $\cP(\Phi_{N+1})$ are
based on the Altarelli-Parisi splitting kernels, weighted by parton
distribution functions (PDFs) where appropriate.

For the specific details of the implementation of the above formulae we refer the reader to Ref.~\cite{Alioli:2015toa}.

The resummed contributions in the previous formulae are obtained from
Soft-Collinear Effective Theory (SCET), where a factorisation formula
for the production of a colour singlet can be written as
\begin{align}\label{eq:factorization}
\frac{\textrm{d} \sigma^{\rm SCET}}{\textrm{d} \Phi_0 \textrm{d}
  \Tau_0}  = & \sum_{ij} H_{ij}(\Phi_0,\mu)\int\! \df r_a\, \df r_b \,B_i(r_a,x_a,\mu) \nn  \\ & \qquad 
B_j(r_b,x_b,\mu) \, S(\Tau_0-\tfrac{r_a+r_b}{Q},\mu)\, .
\end{align}
The sum in the equation above runs over all possible $q\bar{q}$ pairs
$ij=\{ u\bar{u}, \bar{u} u, d \bar{d}, \bar{d} d,\ldots\}$. 
It also depends on the hard $H_{ij}$,
soft $S$ and beam $B_{i,j}$ functions which describe the square of the
hard interaction Wilson coefficients, the soft emissions between
external partons and the hard emissions collinear to the beams
respectively.

In SCET, the resummation of large logarithms is achieved by means of renormalisation group (RG) evolution  between the typical energy scale of each component ($\mu_H, \mu_B $ and $\mu_S$) and a common scale $\mu$.  This
proceeds via convolutions of the single scale factors with the
evolution functions $U_i(\mu_i,\mu)$. The resulting resummed formula for the
$\Tau_0$ spectrum is then given by
\begin{align}
\label{eq:standardresum}
\frac{\textrm{d} \sigma^{\text{NNLL}^\prime}}{\textrm{d} \Phi_0 \textrm{d} \Tau_0} =\;& \sum_{ij} H_{ij}(\Phi_0,\mu_H) \, U_H(\mu_H,\mu)\, \nn \\ & \big\{\big[ B_i(t_a,x_a,\mu_B)\otimes U_B(\mu_B,\mu)\big]\, \nonumber \\
& \times \big[B_j(t_b,x_b,\mu_B)\otimes U_B(\mu_B,\mu)\big] \big\}\, \nn \\ &  \otimes \big[ S(\mu_s)\otimes U_S(\mu_S,\mu)\big]\, ,
\end{align}
where the convolutions between the different functions are written in a
schematic form.  In order to reach NNLL$^\prime$ accuracy, we need to know
the boundary conditions of the evolution, namely the hard, beam and
soft functions up to NNLO accuracy, and the cusp (non-cusp) anomalous
dimensions up to three-(two-)loop order. 
The  beam and
soft functions at NNLO accuracy are available in the literature~\cite{Kelley:2011ng,Monni:2011gb,Gaunt:2014xga}, as well as the cusp (non-cusp) anomalous
dimensions up to three-(two-)loop order~\cite{Idilbi:2006dg,Becher:2006mr,Hornig:2011iu,Kang:2015moa,Gaunt:2015pea}.

\begin{figure*}[ht!]
  \begin{subfigure}[b]{\rescalethreeplots}
    \includegraphics[width=\textwidth]{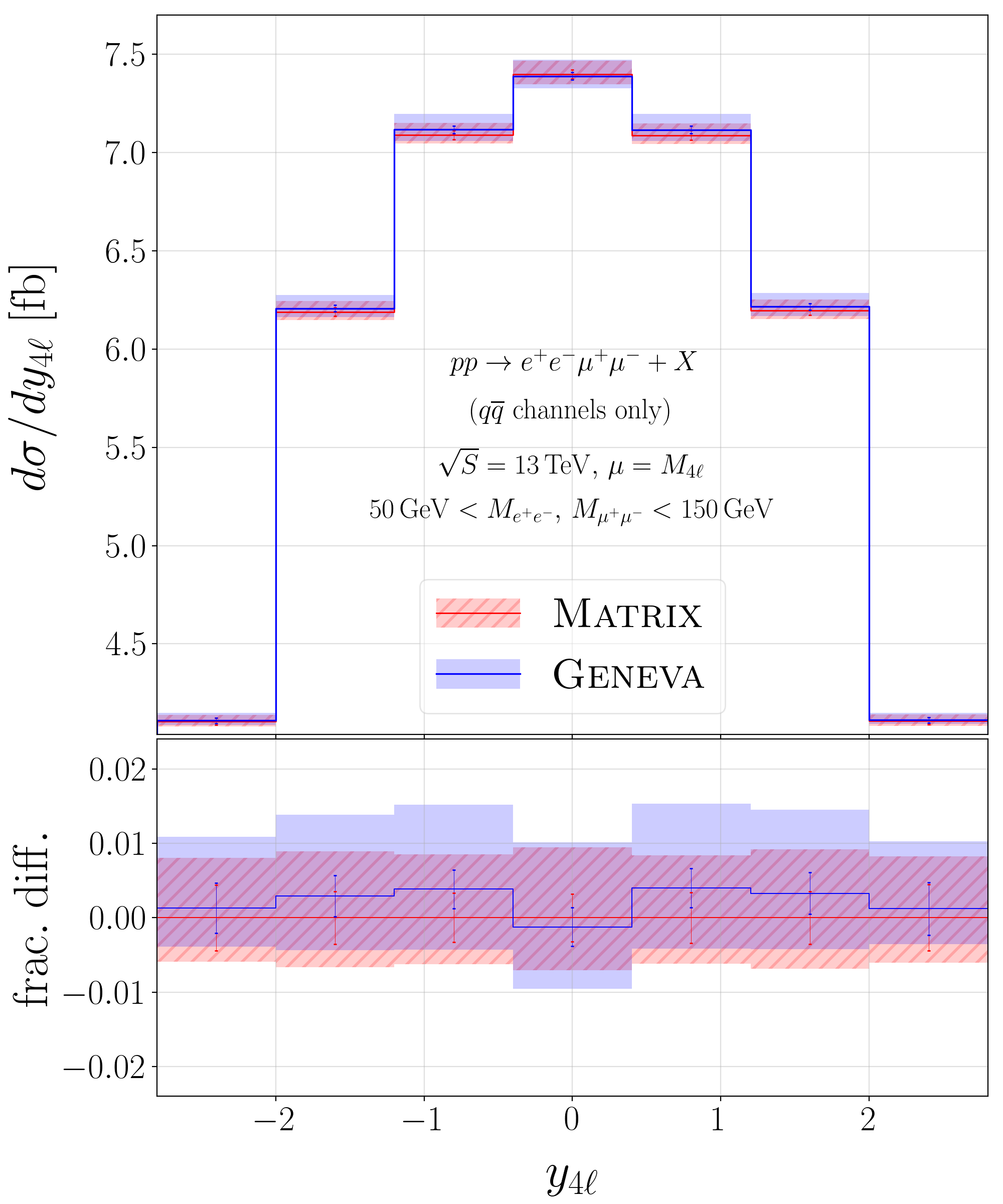}%
  \end{subfigure}
  \hspace*{\hspacebetweenthreeplots}
  \begin{subfigure}[b]{\rescalethreeplots}
    \includegraphics[width=\textwidth]{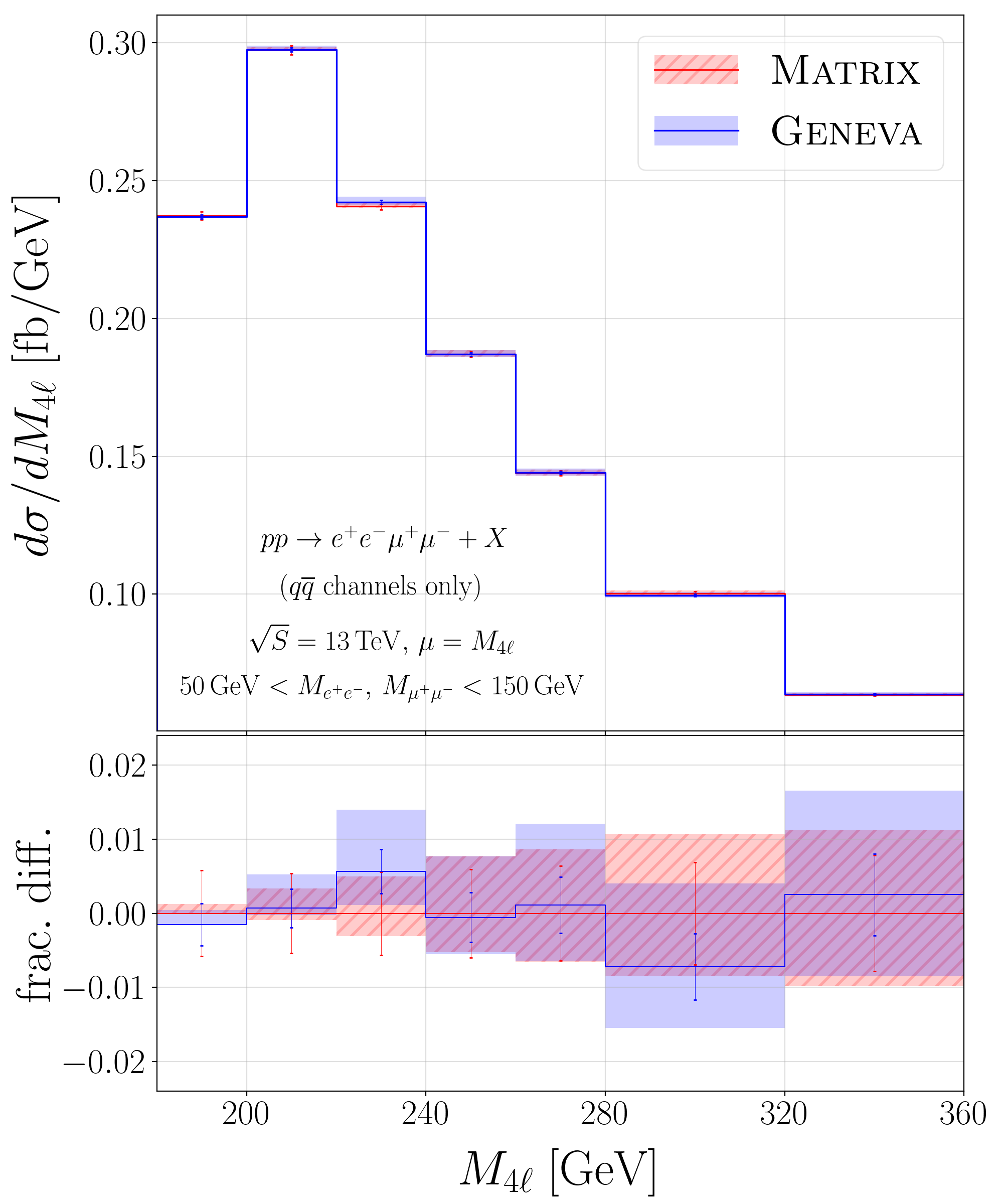}%
  \end{subfigure}
  \hspace*{\hspacebetweenthreeplots}
  \begin{subfigure}[b]{\rescalethreeplots}
    \includegraphics[width=\textwidth]{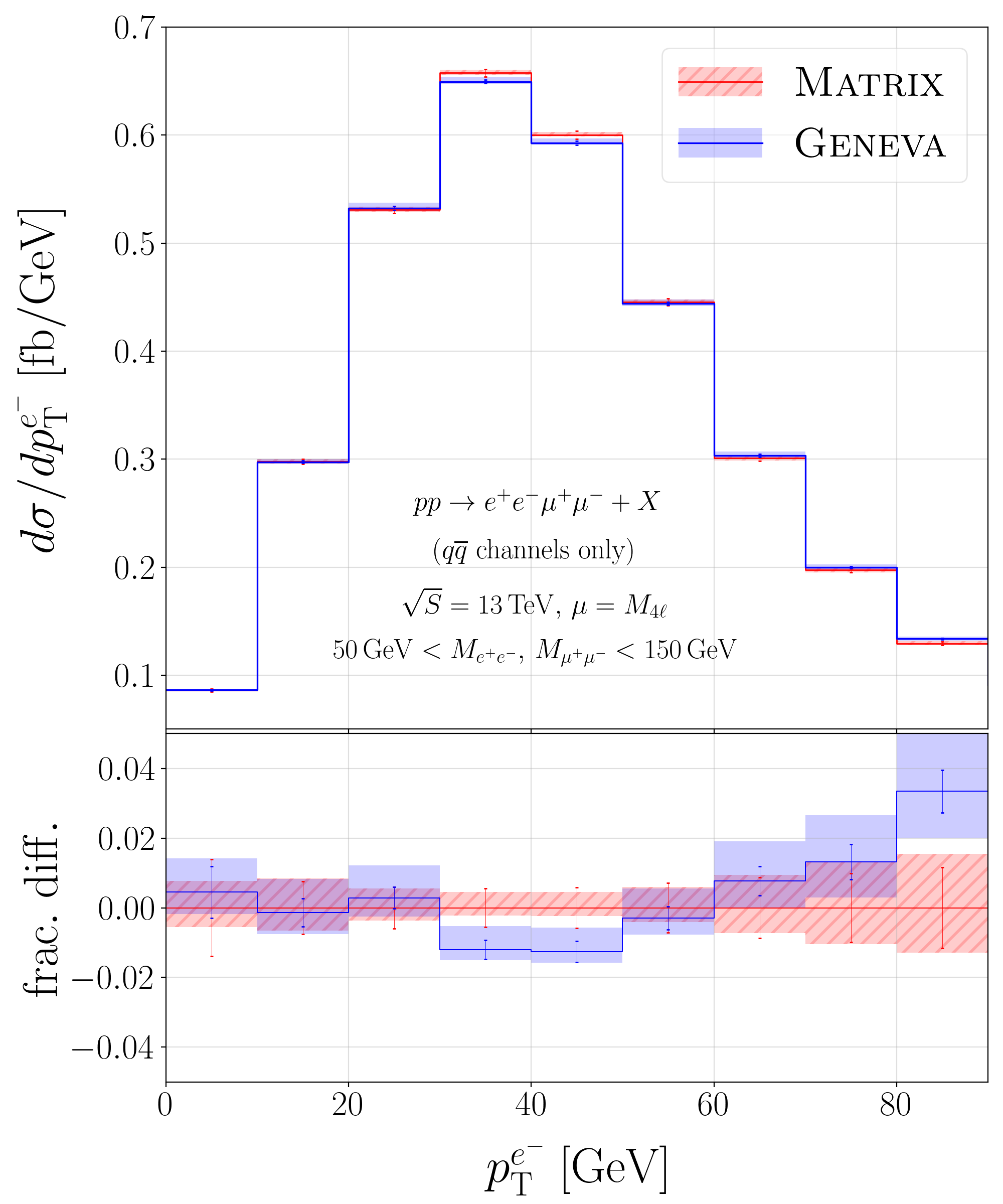}%
  \end{subfigure}
  \vspace{\spaceabovefigurecaption}
  \caption{Comparison between the \matrix and \geneva predictions at NNLO accuracy. We show the
    rapidity of the four leptons (left), the mass of the four
    leptons (centre) and the
    transverse momentum of the electron (right). Scale uncertainty bands include 3-point renormalisation and factorisation scale variations. Statistical errors connected to the Monte Carlo integration are shown as vertical error bars.
    \label{fig:gvavsmatrixqqbar}
  }
\end{figure*}

The two-loop hard function $H_{ij}^{(2)}$ was
instead computed starting from the form factors calculated in
Refs.~\cite{Caola:2014iua,Gehrmann:2015ora} and using the numerical implementation in the public code \textsc{VVamp}~\cite{vvamp}.
Finally, we use  \openloopsTwo
\cite{Buccioni:2019sur,Cascioli:2011va,Buccioni:2017yxi} for the calculation of all the remaining  matrix elements.

Starting from these accurate parton-level predictions we can interface
to the \pythiaEight parton shower to produce the high-multiplicity
final states that can in turn be compared to experimental data.  The
shower adds extra radiation to the exclusive $0$- and $1$-jet cross
sections and extends the inclusive $2$-jet cross section by including
higher jet multiplicities. This means that we can expect the shower to
modify the distributions of exclusive observables sensitive to the
radiation, preserving the leading logarithmic accuracy for observables other
than $\Tau_0$. At the same time, we have verified that the shower does
not modify any distribution which is inclusive over the radiation, as
was the case for the \geneva implementation of similar colour-singlet
production processes. As a consequence we maintain  NNLO accuracy
for inclusive observables.

All the details about the interface between \geneva and
\pythiaEight can be found in section 3 of \cite{Alioli:2015toa}.

\section{Results and comparison to LHC data}
\label{sec:results}

In the following, we focus on the process
\begin{equation}
  p p \rightarrow e^+ e^- \mu^+ \mu^- + X
\end{equation}
at a hadronic centre-of-mass energy of 13 $\TeV$ and require that
the masses of both lepton--antilepton pairs are between 50 and 150
$\GeV$. We use the PDF set \verb|NNPDF31_nnlo_as_0118|~\cite{Ball:2017nwa} from LHAPDF6
\cite{Buckley:2014ana} and set both the renormalisation and
factorisation scales to the mass $M_{4\ell}$ of the four-lepton
system. We choose the resolution cutoffs to be  $\Tau_0^\cut = 1$ GeV and $\Tau_1^\cut = 1$ GeV. 

For the validation, we focus only on the quark--antiquark
channels and neglect the  loop-induced gluon fusion channel, which only
starts appearing in the calculation at NNLO and can therefore be added
as a nonsingular fixed-order contribution. At this energy, the latter
contribution amounts
to $\sim 6$\% of the total cross section and,  as such, its inclusion will be
important when comparing to data.

In \fig{gvavsmatrixqqbar} we validate the NNLO accuracy of the \geneva
results by comparing with those obtained via an independent NNLO
calculation implemented in \matrix. In particular, we show the
distributions of the rapidity $y_{4\ell}$, mass $M_{4\ell}$ of the
four leptons and the transverse momentum $p_\T^{e^-}$ of the
electron. We observe a good agreement, with the only differences
appearing in the shape of the $p_\T^{e^-}$ distribution. This is
likely to be a consequence of the additional higher-order effects
provided by \geneva, as observed in previous \geneva predictions for
other colour-singlet production processes. We observe, however, that in
almost all the bins the theoretical uncertainty bands computed by
\geneva and \matrix still overlap.

Next we turn on the shower effects by interfacing to \pythiaEight. In
order to maintain a simple analysis routine and focus only on QCD
corrections, we do not include QED effects and multiple-particle
interactions (MPI) in the shower.

\begin{figure}[ht!]
    \includegraphics[width=0.45\textwidth]{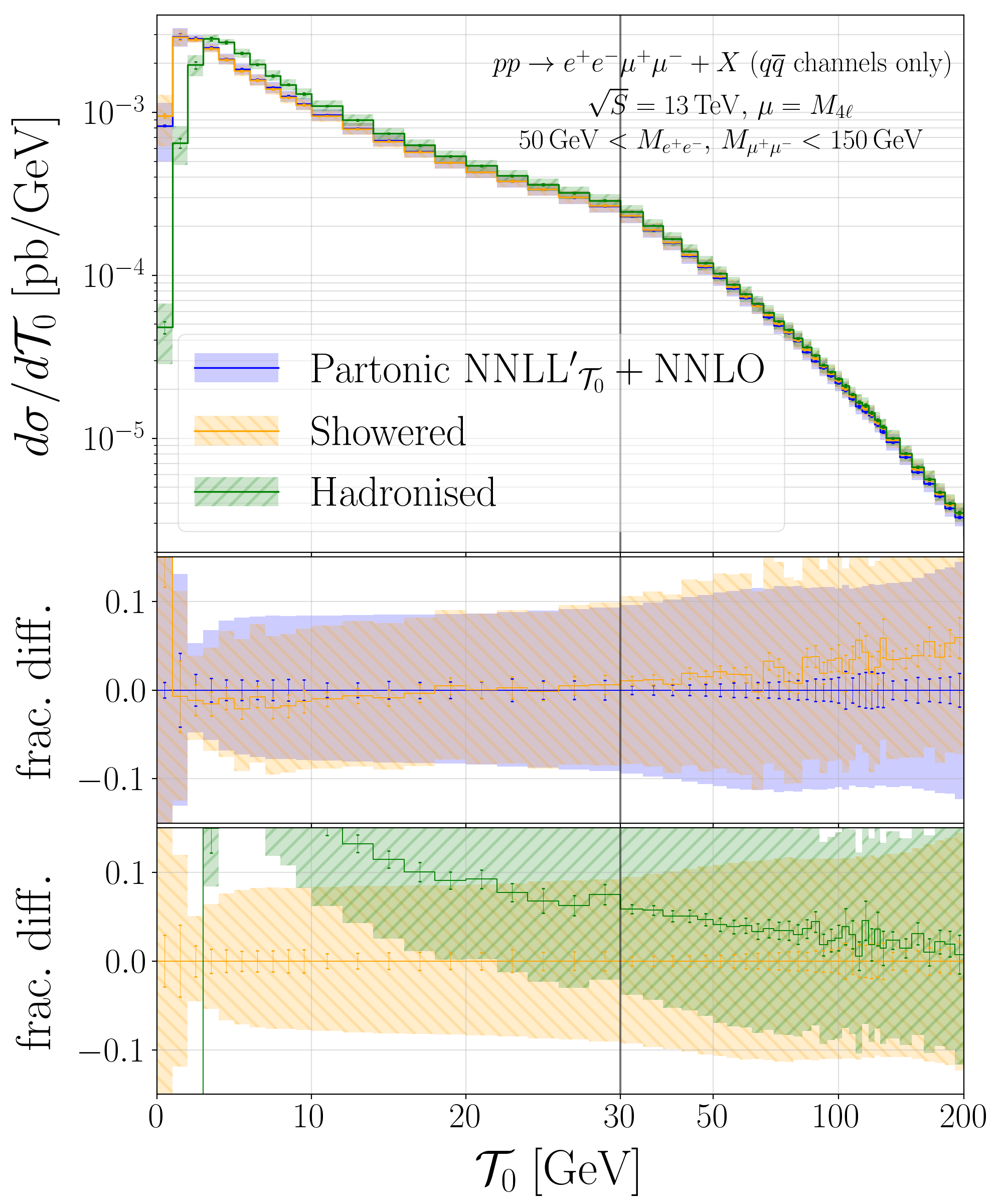}%
  \vspace{\spaceabovefigurecaption}
  \caption{Comparison between the partonic, showered and hadronised results for the 
   beam-thrust $\mathcal{T}_0$ distribution in \geneva and \geneva+\pythiaEight.
    \label{fig:partonic_showered_hadronised}
  }
\end{figure}

In \fig{partonic_showered_hadronised} we present our predictions for
the NNLL$^\prime_{\Tau_0}$+NNLO beam-thrust spectrum (partonic result)
and study the effect of shower and hadronisation on the $\Tau_0$
distribution. In order to highlight the peak and transition regions
where the resummation effects are more important, we show the results
on a semi-logarithmic scale, which is linear up to the value of
$30$~GeV and logarithmic beyond. In the first ratio plot, we compare the $\Tau_0$
distribution before and after the shower, observing that the
NNLL$^\prime_{\Tau_0}$+NNLO accuracy reached at parton level is
numerically very well preserved. The largest difference is, as
expected, in the first bin where the shower generates all the events
with $\Tau_0<\Tau_0^\cut$ which were previously integrated out in the
inclusive $0$-jet cross section. In the second ratio plot we instead
show the effect of hadronisation on the showered distribution. Owing
to its nonperturbative origin, hadronisation affects mostly the region
of small $\Tau_0$ and becomes less and less important in the tail of
the distribution.

\begin{figure*}[ht!]
  \begin{subfigure}[b]{\rescalethreeplots}
    \includegraphics[width=\textwidth]{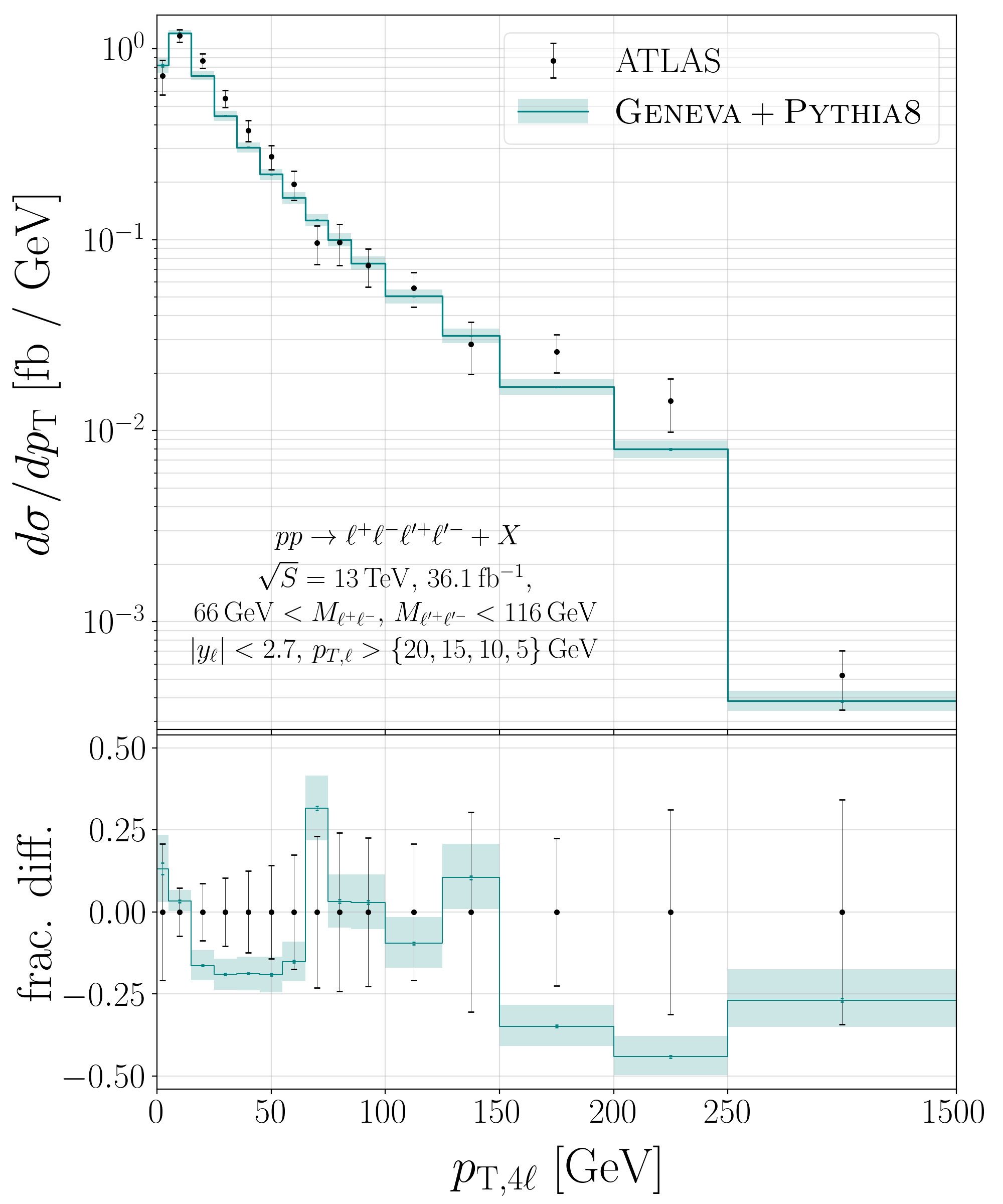}%
  \end{subfigure}
  \hspace*{\hspacebetweenthreeplots}
  \begin{subfigure}[b]{\rescalethreeplots}
    \includegraphics[width=\textwidth]{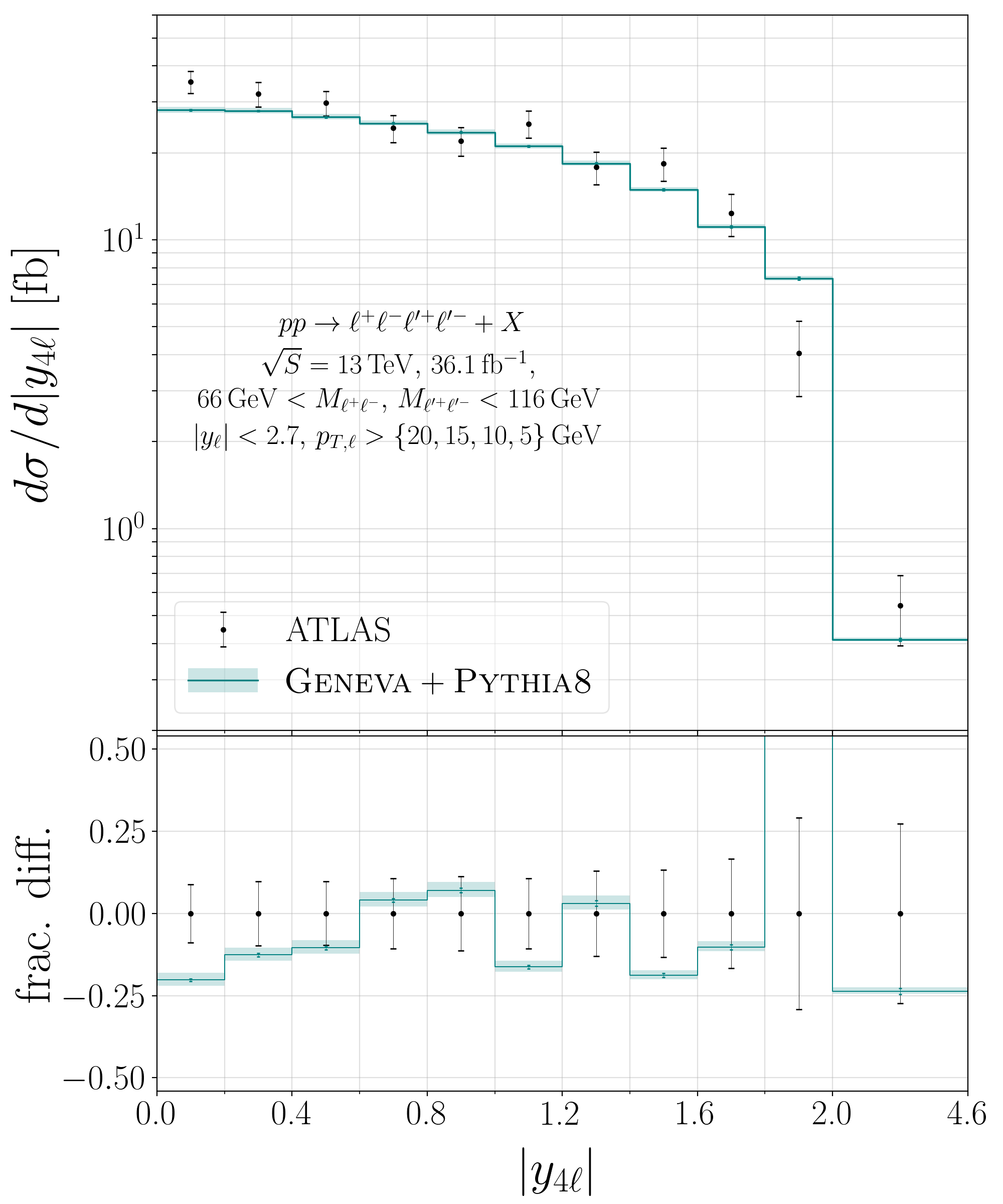}%
  \end{subfigure}
  \hspace*{\hspacebetweenthreeplots}
  \begin{subfigure}[b]{\rescalethreeplots}
    \includegraphics[width=\textwidth]{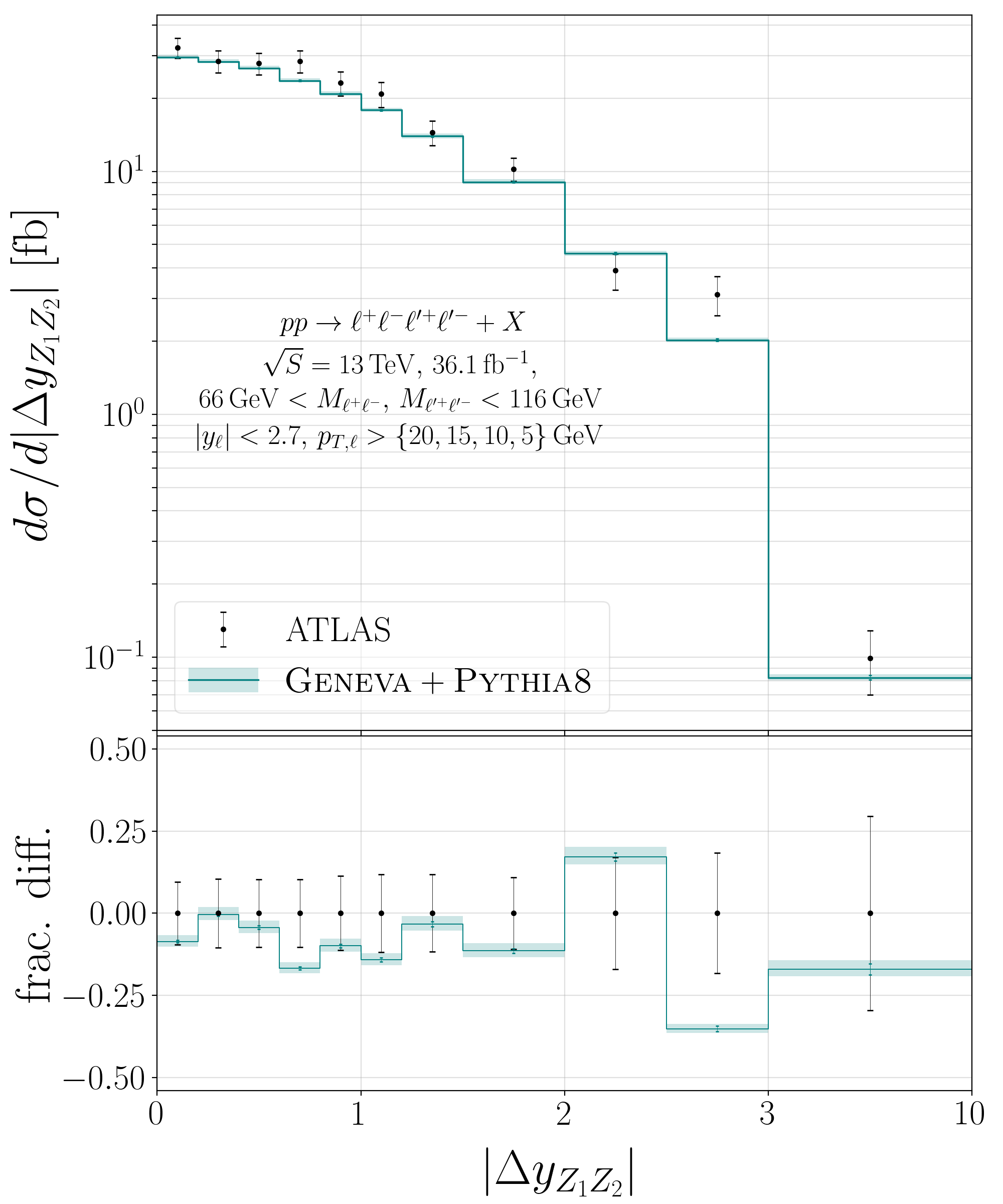}%
  \end{subfigure}
  \begin{subfigure}[b]{\rescalethreeplots}
    \includegraphics[width=\textwidth]{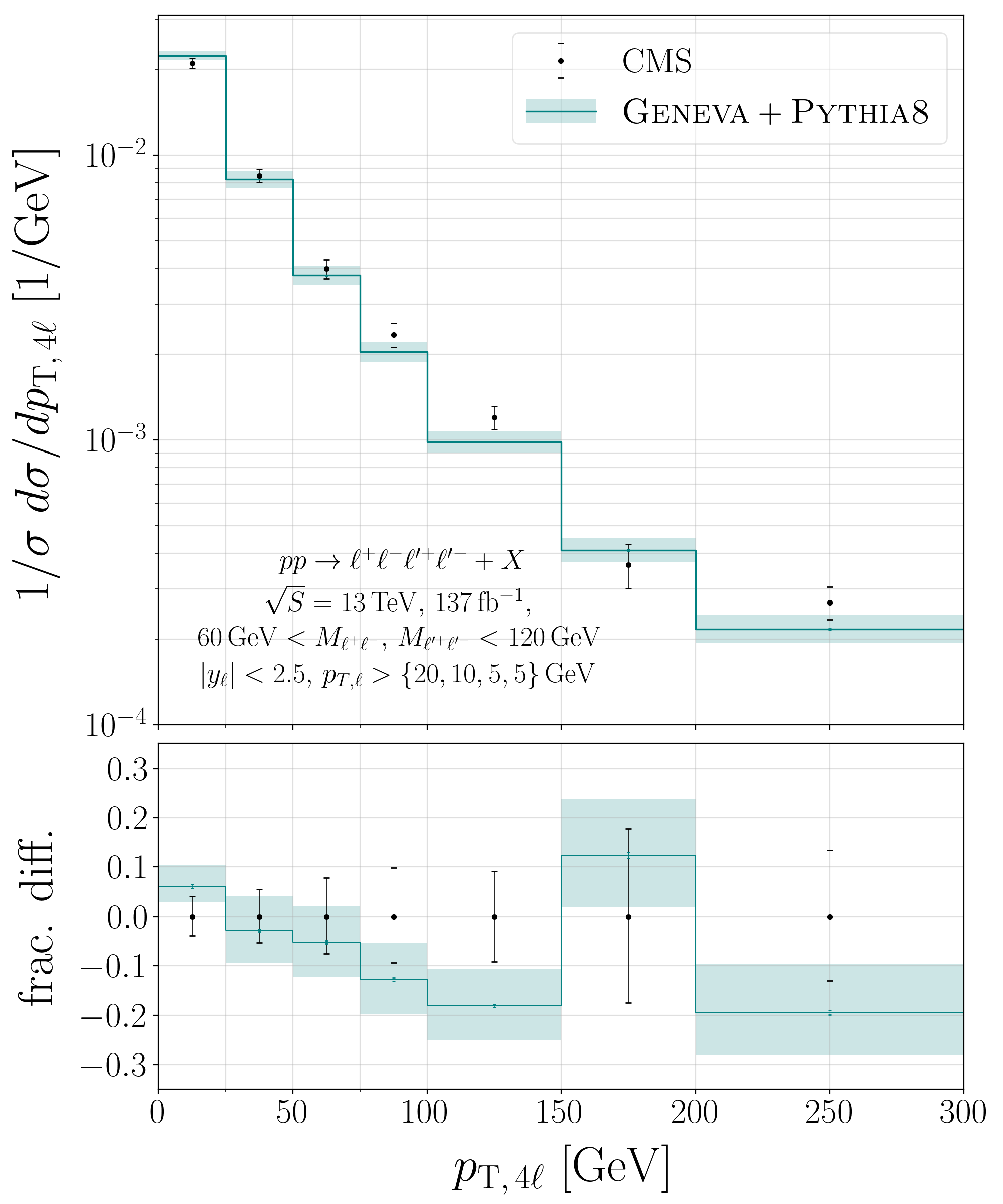}%
  \end{subfigure}
  \hspace*{\hspacebetweenthreeplots}
  \begin{subfigure}[b]{\rescalethreeplots}
    \includegraphics[width=\textwidth]{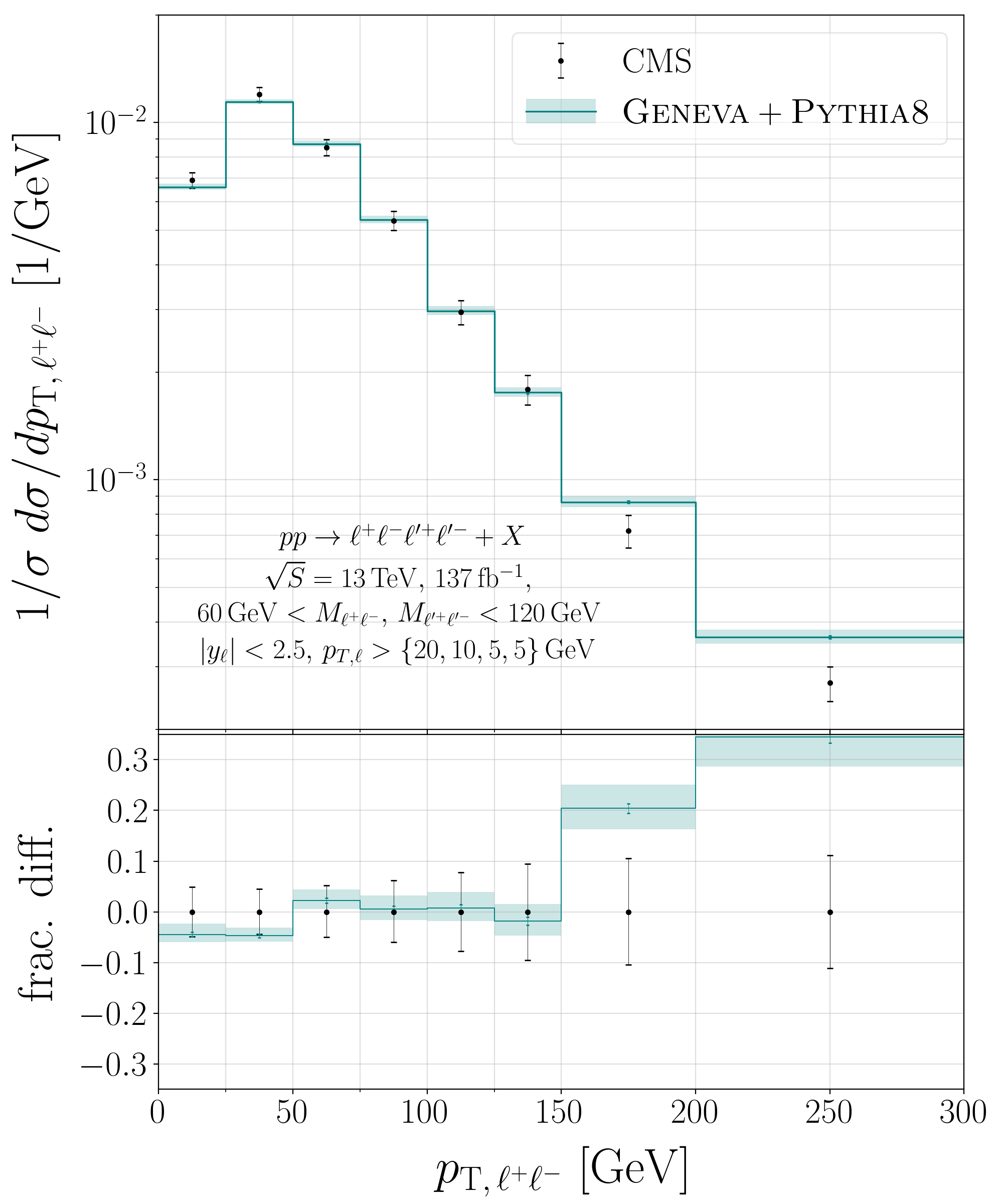}%
  \end{subfigure}
  \hspace*{\hspacebetweenthreeplots}
  \begin{subfigure}[b]{\rescalethreeplots}
    \includegraphics[width=\textwidth]{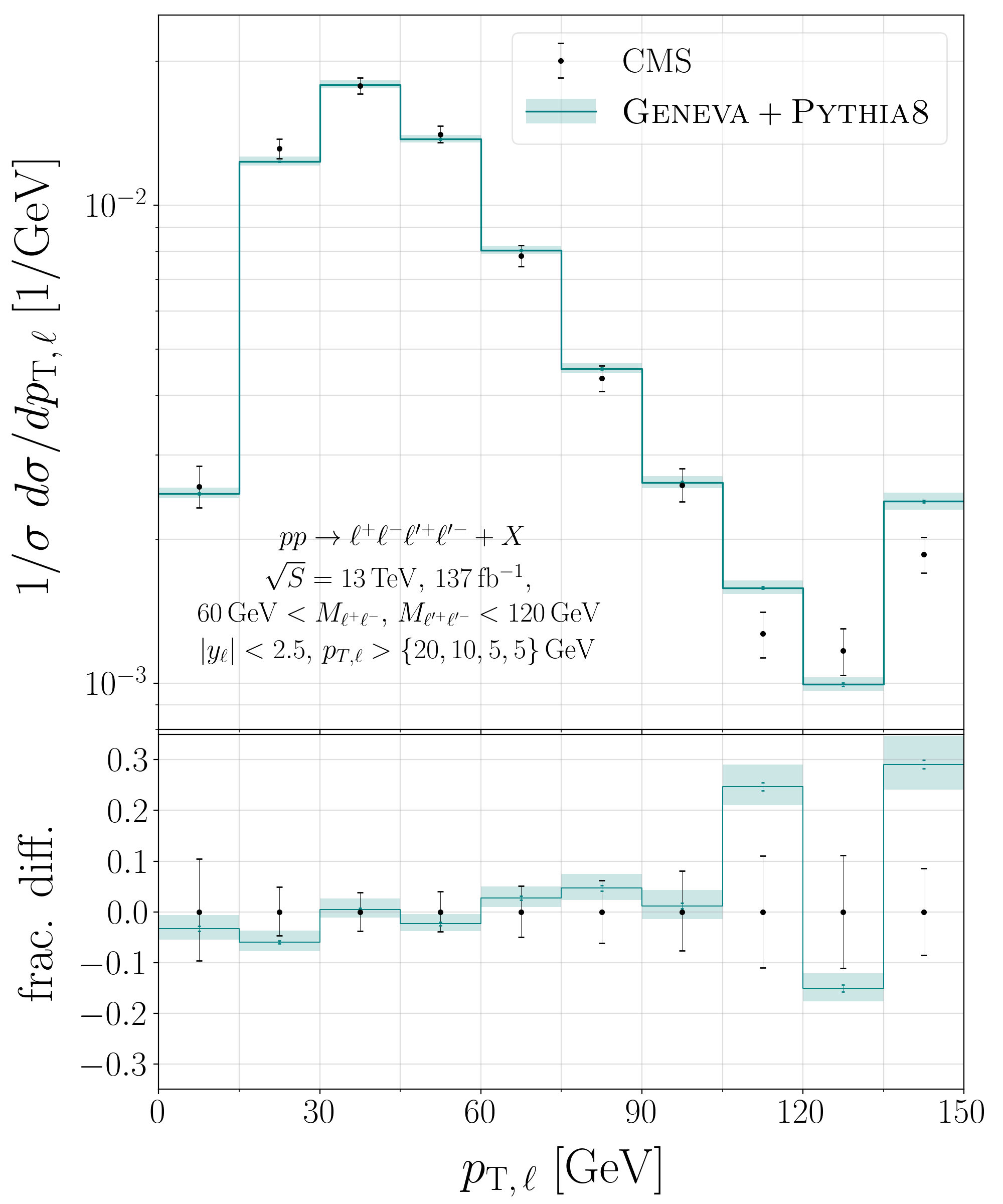}%
  \end{subfigure}
  \vspace{\spaceabovefigurecaption}
  \caption{Comparison to ATLAS and CMS measurements from the LHC at 13 TeV. Selection cuts, bin widths and observable definitions are as detailed in the original ATLAS~\cite{Aaboud:2017rwm} and CMS~\cite{Sirunyan:2020pub} publications.}
  \label{fig:datacomparison}
\end{figure*}

Finally, in \fig{datacomparison} we compare our predictions with data obtained at the LHC at
13 $\TeV$ both from the ATLAS~\cite{Aaboud:2017rwm} and CMS~\cite{Sirunyan:2020pub} experiments.
For this comparison  we now include the loop-induced gluon fusion  channel and MPI effects.
The binning of  observables and  the acceptance cuts applied to the events in
each analysis have been modelled according to the experimental requirements.
They are briefly listed in the figure, but we refer the reader to the original papers for all the  details.

In the first row we show the comparison with ATLAS data, obtained with an integrated luminosity of $36.1 $ fb$^{-1}$. We consider the
distributions for the transverse momentum of the four leptons
$p_{\T,4\ell}$, and two more  inclusive
distributions,  
the absolute value $|y_{4\ell}|$ of the rapidity of the four leptons
and the absolute value $|\Delta y_{Z_1 Z_2}|$ of the difference in
rapidity between the two reconstructed $Z$ bosons.
The agreement with data is reasonably good but the reduced experimental statistics make it difficult to draw more precise
conclusions. We only notice a possible tension in
the distribution of $|y_{4\ell}|$, where for small values of the
rapidity the data seem to be systematically above the \geneva
predictions. This behaviour has, however, been observed previously with
several other Monte Carlo event generators in the original ATLAS publication~\cite{Aaboud:2017rwm}.

In the second row, we show the comparison with CMS data, obtained with an integrated luminosity of $137 $ fb$^{-1}$. We focus on the normalised distributions for $p_{\T,4\ell}$, the transverse momentum $p_{\T,\ell^+\ell^-}$ of
the vector bosons and the transverse momentum $p_{\T,\ell}$ of the
leptons.
The latter two distributions are obtained by averaging over all vector bosons and leptons present in an event.
Due to the increased luminosity and the fact that we are comparing normalised distributions, here we observe smaller experimental uncertainties  and
a better agreement between the data and the \geneva
predictions. We highlight the fact that according to the original definition of bins in Ref.~\cite{Sirunyan:2020pub}, the last bin of each distribution we show must be interpreted as an overflow bin containing the
contributions from the lower border up to the maximum possible value
for that observable. This justifies for example the peculiar
behaviour of the last bin of the $p_{\T,\ell}$ distribution.

We conclude by noticing that the transverse momentum
$p_{\T,\ell^+\ell^-}$ of the vector bosons shows a sizable difference
for values larger than $150$~GeV, where EW effects are known to be important~\cite{Kallweit:2019zez}
 and should thus be included to improve the agreement with data.

\section{Conclusions}
\label{sec:conc}

In this Letter, we have presented the first event generator for  the
production of a pair of $Z$ bosons decaying into four leptons at NNLO accuracy matched to
the \pythia8 parton shower.

This was obtained using the resummation  of the $0$-jettiness $\Tau_0$
resolution variable at NNLL$^\prime$ accuracy,  matched to a fixed-order
calculation at NNLO precision. The calculation was   performed within the \geneva
Monte Carlo framework, which allowed us to interface to the \pythiaEight  parton
shower and hadronisation models.

After successfully validating the NNLO accuracy of the results against
\matrix, we studied the effect of the shower on the differential distributions,
observing that they are numerically small for inclusive quantities, as expected.
Consequently, their NNLO accuracy is correctly  maintained.
We also verified that the
hadronisation has a significant impact only for exclusive observables in the region of small
$\Tau_0$, where the nonperturbative effects are not negligible.

Finally, we compared to LHC data, both from the ATLAS and CMS
experiments. We found good agreement, except for
a tension observed in
the region of small absolute rapidity of the four lepton system.  A
similar behaviour had already been observed for other Monte Carlo
generators. We also observed a general overestimation of the production rate for
vector bosons at large transverse momentum, motivating the need for the inclusion of EW corrections
to improve the agreement with data in this region.

Possible future directions for improvement for this calculation would be the
inclusion of the NLO QCD corrections to the gluon fusion channel and of the 
aforementioned  NLO EW corrections.

The code used for the simulations presented in this work is available
upon request from the authors and will be made public in a future
release of \geneva.

\section*{Acknowledgements}
\label{sec:Acknowledgements}
SA is grateful to Pietro Govoni and Raquel Gomez Ambrosio for discussions and
for their help with the CMS analysis.  The work of SA, AB, AG, SK, MAL, RN, and DN
is supported by the ERC Starting Grant REINVENT-714788. SA also acknowledges
funding from Fondazione Cariplo and Regione Lombardia, grant 2017-2070. The
work of SA is also supported by MIUR through the FARE grant R18ZRBEAFC. We
acknowledge the CINECA award under the ISCRA initiative and the National
Energy Research Scientific Computing Center (NERSC), a U.S. Department of
Energy Office of Science User Facility operated under Contract
No. DEAC02-05CH11231, for the availability of the high performance computing
resources needed for this work.

\appendix

\section{Nonsingular power corrections}
\label{app:power_corrections}

\begin{figure}[ht!]
    \includegraphics[width=0.45\textwidth]{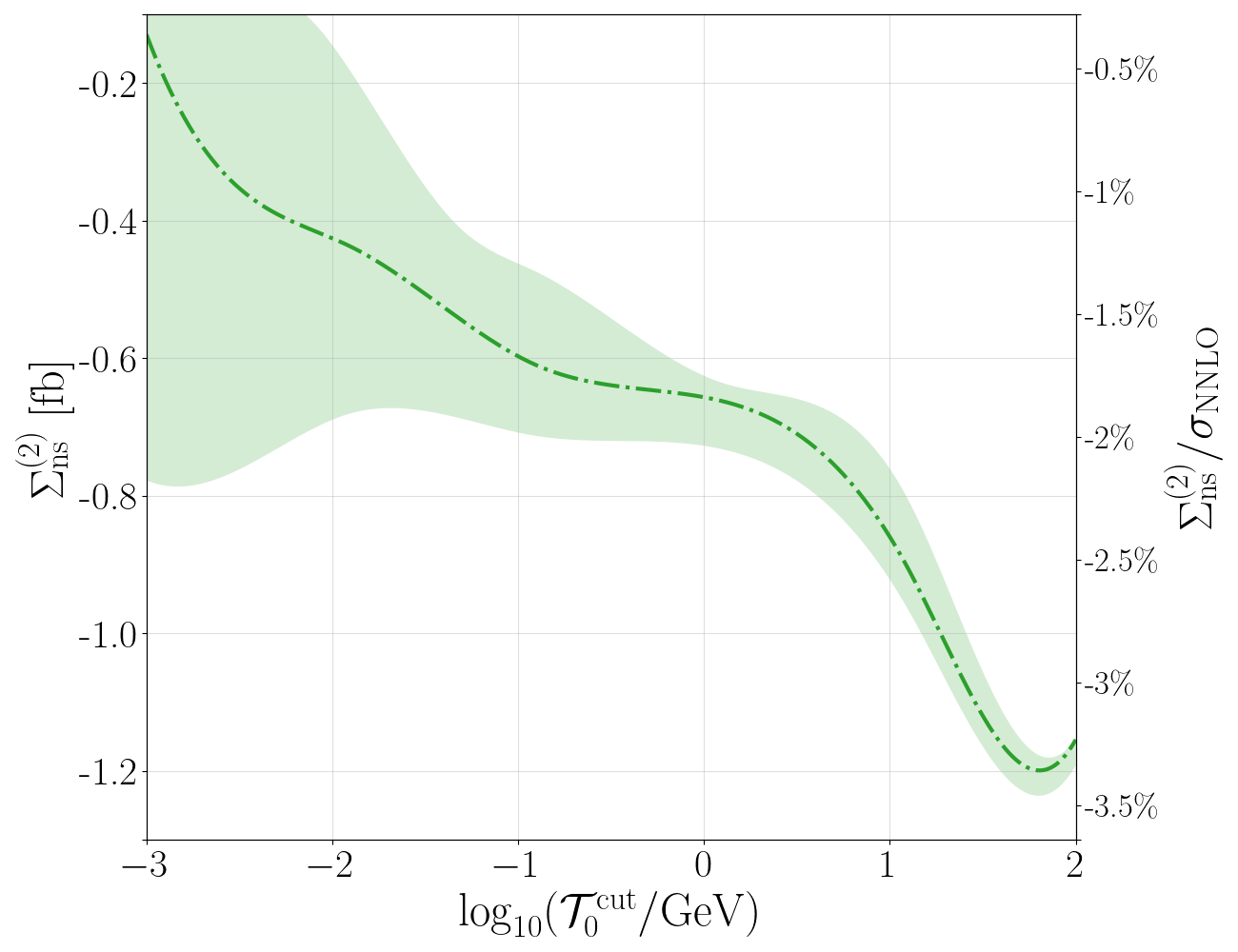}%
  \vspace{\spaceabovefigurecaption}
  \caption{Size of the neglected power corrections
    $\Sigma^{(2)}_{\mathrm{ns}}$ as a function of the resolution
    parameter $\Tau_0^\cut$. The target result $\sigma_{\rm NNLO}$ is
    the fixed-order cross section, as computed by \matrix.
    \label{fig:power_corrections}
  }
\end{figure}

The \geneva calculation is based on $N$-jettiness subtraction, using a resolution cutoff $\Tau_0^\cut$. The
contributions below the cut, given in \eq{0masterful},
require a local NNLO subtraction for their implementation.
In \geneva, exploiting the $N$-jettiness subtraction,  we substitute the expression given in \eq{0masterful} with
\begin{align} \label{eq:0masterfultilde}
\frac{\dsigtildeMCz}{\df\Phi_0}(\Tau_0^\cut)
&= \frac{\df\sigma^{\rm NNLL'}}{\df\Phi_0}(\Tau_0^\cut)
- \biggl[\frac{\df\sigma^{\rm NNLL'}}{\df\Phi_0}(\Tau_0^\cut) \biggr]_{\rm NLO_0}
\nn \\ & \quad
+ (B_0 + V_0)(\Phi_0)
\nn \\ & \quad
+ \int \! \frac{\df \Phi_1}{\df \Phi_0}\, B_1(\Phi_1)\, \theta[\Tau_0(\Phi_1) < \Tau_0^\cut]
\,.
\end{align}
In other words, we neglect the contribution
\begin{align} \label{eq:nonsingcum}
\frac{d\Sigma^{(2)}_{\mathrm{ns}}}{\df\Phi_0}(\Tau_0^\cut)
&= \frac{\dsigMC_0}{\df\Phi_0}(\Tau_0^\cut) - \frac{\dsigtildeMCz}{\df\Phi_0}(\Tau_0^\cut)
\nn \\
&= \biggl[\frac{\df\sigma^{\rm NNLL'}}{\df\Phi_0}(\Tau_0^\cut)
  \biggr]_{\rm NLO_0}
\nn \\ & \quad
- \biggl[\frac{\df\sigma^{\rm NNLL'}}{\df\Phi_0}(\Tau_0^\cut) \biggr]_{\rm NNLO_0}
\nn \\ & \quad
+ W_0(\Phi_0)
\nn \\ & \quad
+ \int \! \frac{\df \Phi_1}{\df \Phi_0}\, V_1(\Phi_1)\, \theta[\Tau_0(\Phi_1) < \Tau_0^\cut]
\nn \\ & \quad
+ \int \! \frac{\df \Phi_2}{\df \Phi_0}\, B_2 (\Phi_2)\, \theta[\Tau_0(\Phi_2) < \Tau_0^\cut]
\,.
\end{align}
These terms that we are neglecting are power corrections of $\ord{\alpha_s^2}$
 -- their integral is shown in \fig{power_corrections} as a
function of $\Tau_0^\cut$. On the right axis we also show the relative size
of this corrections as a fraction of the NNLO cross section $\sigma_{\rm
  NNLO}$, computed by \matrix. As expected, we observe
that the difference between the \geneva and the \matrix results
becomes smaller as the value of $\Tau_0^\cut$
decreases. At extremely small values of $\Tau_0^\cut$, however, the computation becomes numerically unstable and subject to large statistical uncertainties,
reflected in the increased statistical error band. For the results presented in this Letter we set
\begin{equation}
  \Tau_0^\cut = 1 \GeV\,.
\end{equation}
Since these missing contributions only affect the events below the $\Tau_0^\cut$, we can recover the correct NNLO cross section by reweighting  them by the difference. As for any event generator which requires an infrared-safe definition of events at higher orders, we  miss the nonsingular kinematical dependence at $\ord{\alpha_s^2}$  for events below the cut.
We validated that this  does not produce significant distortions in differential distributions by comparing them to \matrix in \fig{gvavsmatrixqqbar}.

\begin{figure}[ht!]
    \includegraphics[width=0.45\textwidth]{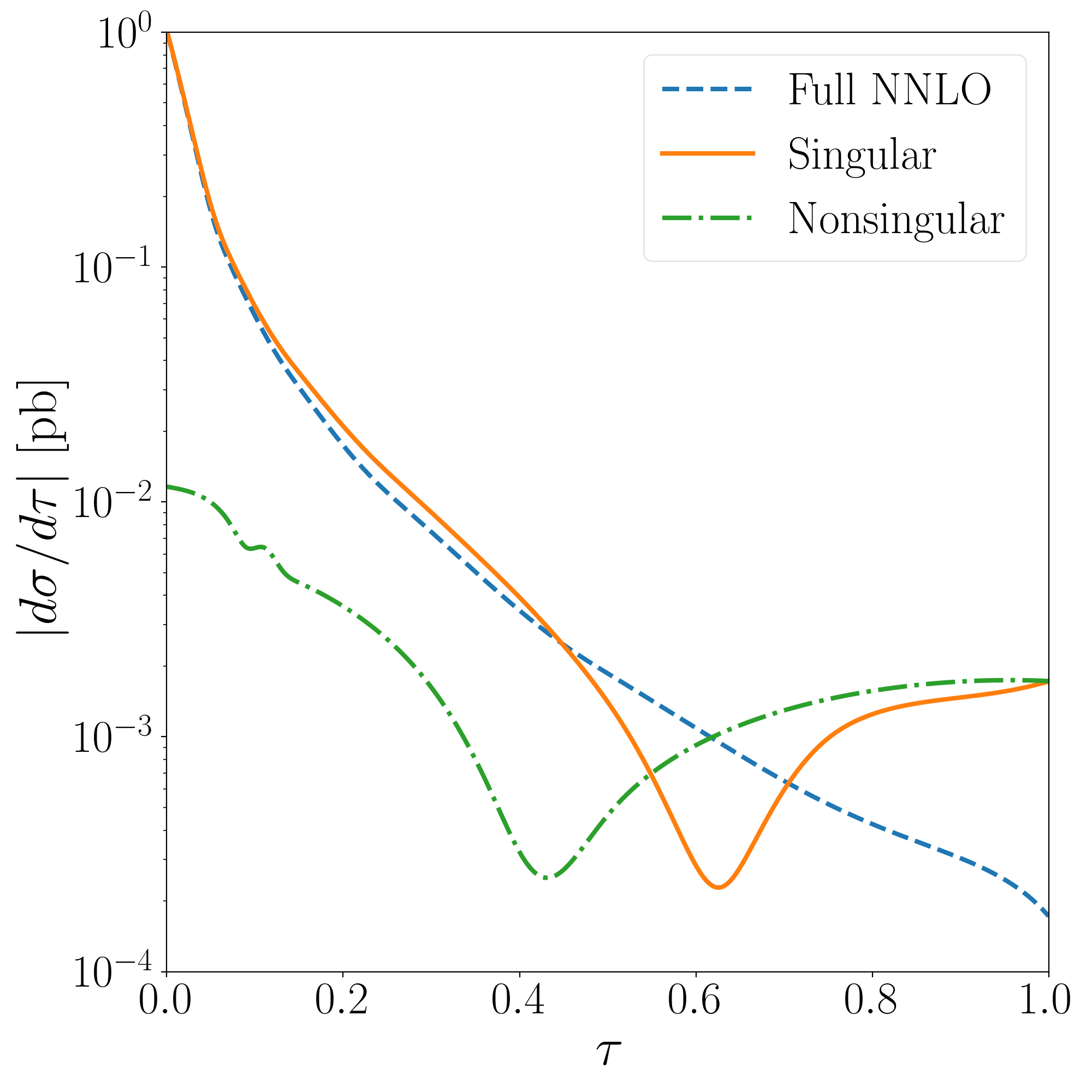}%
  \vspace{\spaceabovefigurecaption}
  \caption{Comparison between the fixed-order distribution (Full
    NNLO), the absolute value of the expansion of the resummed
    contribution up to order $\alpha_S^2$ (Singular) and the absolute
    value of their difference (Nonsingular), as a function of $\tau$.
    \label{fig:profile_functions}
  }
\end{figure}

\section{Profile functions}
\label{app:profile_functions}

The scales $\mu_H$, $\mu_B$, and $\mu_S$ at which the hard, beam and
soft functions are evaluated are chosen using profile functions~\cite{Stewart:2013faa}, following the prescriptions
illustrated in Ref.~\cite{Alioli:2020qrd}. Here we limit ourselves to
explaining briefly the meaning of the four parameters on which they
depend, namely $y_0$, $x_1$, $x_2$, and $x_4$, and how we set
them. According to those prescriptions, the hard scale is kept fixed
to $\mu_H = Q = M_{4\ell}$, while the beam and soft scales are run
to resum the large logarithms. If we define a dimensionless
\begin{equation}
  \tau = \frac{\Tau_0}{Q}\,,
\end{equation}
in the region where $2y_0/Q < \tau < x_1$, the scales are chosen
canonically, namely
\begin{equation}
  \mu_B = \sqrt{\Tau_0Q} \qquad \mu_S = \Tau_0\,.
\end{equation}
For $\tau < 2y_0/Q$ the beam and soft scales are modified in a
continuous way, so that, for $\tau \rightarrow 0$,
\begin{equation}
  \mu_B \rightarrow \sqrt{y_0Q} \qquad \mu_S \rightarrow y_0\,.
\end{equation}
This prevents the scales from reaching nonperturbative values, where
the resummation framework breaks down. For this process we maintain the choice made in Ref.~\cite{Alioli:2020qrd}
and set
\begin{equation}
  y_0 = 2.5 \GeV\,.
\end{equation}
At the opposite end of the spectrum, for  $\tau > x_3$ the resummation must be switched off in order to recover the fixed-order result. This is achieved by setting
\begin{equation}
  \mu_B = \mu_S = Q\,.
\end{equation}
The parameter $x_2$ is used to choose how we turn off the
resummation, usually picking it as the middle
point between $x_1$ and $x_3$. In order to set the values for $x_1$
and $x_3$ we look at \fig{profile_functions}, where we compare the
fixed-order distribution (Full NNLO), the absolute value of the
expansion of the resummed contribution up to  $\ord{\alpha_S^2}$
(Singular) and the absolute value of their difference
(Nonsingular). We observe that the first two curves almost overlap in
the region where $\tau \lesssim 0.2$ and that the nonsingular
contribution becomes important  at $\tau \approx
0.6$. Based on these considerations we choose the values
\begin{equation}
\{x_1, x_2, x_3\} = \{0.2, 0.4, 0.6\}\,.
\end{equation}


\bibliographystyle{elsarticle-num}
\bibliography{geneva}

\begin{thebibliography}{10}
\bibitem{munich}
{\sc Munich} is the abbreviation of ``MUlti-chaNnel Integrator at Swiss (CH)
  precision''---an automated parton level NLO generator by S.~Kallweit. In
  preparation.
\end{thebibliography}

\end{document}